\documentclass[11pt]{article}
\usepackage{graphics}
\usepackage{epsfig}
\usepackage{rotating}
\newcommand{\etal}{et al.}

\newcommand{\eg}{e.g. }
\newcommand{\ie}{i.e. }
\newcommand{\etc}{etc.}
\newcommand{\ltsim}{\raisebox{-1mm}{$\stackrel{<}{\sim}$}}

\def\arcm{\hbox{$^\prime$}}
\def\arcs{\arcm\hskip -0.1em\arcm}

\begin{document}                        	

\begin{center}

\bf{\Large Analysis of the XMM-Newton EPIC Background: Production of
Background Maps and Event Files} \\
\vspace{3mm} 
\bf{\large Andrew M.\,Read} \\ 
\vspace{3mm} 
\bf{\large School of Physics \& Astronomy, University of Birmingham,} \\
\bf{\large Birmingham B15 2TT, UK} \\
{\em amr@star.sr.bham.ac.uk}\hspace{5mm} \verb£http://www.sr.bham.ac.uk/~amr£ \\
\verb£http://www.sr.bham.ac.uk/xmm3/£

\end{center}

\section*{Abstract}

XMM-Newton background maps for the 3 EPIC instruments in their
different instrument/mode/filter combinations and in several energy
bands have been constructed using a superposition of many pointed
observations. Event datasets for the different instrument/mode/filter
combinations have also been constructed, with longer exposure times
than previously created files. This document describes the
construction of the background maps and event files, and their usage.
Further details on how to obtain these background products, together
with related software and these explanatory notes can be found at
\verb£http://www.sr.bham.ac.uk/xmm3/BGproducts.html£.

\section{Introduction} 

The XMM-Newton observatory provides unrivalled capabilities for
detecting low surface brightness emission features from extended and
diffuse galactic and extragalactic sources, by virtue of the large
field of view of the X-ray telescopes with the EPIC MOS (Turner \etal\
2001) and pn (Str\"{u}der \etal\ 2001) cameras at the foci, and the
high throughput yielded by the heavily nested telescope mirrors. The
satellite has the largest collecting area of any imaging X-ray
telescope.

In order to exploit the excellent EPIC data from extended objects, the
EPIC background, known now to be higher than estimated pre-launch,
needs to be understood thoroughly. With a good model of the particle
and photon background, one can correctly background-subtract images
and spectra extracted over different energy bands and from different
areas of the detectors.

The initial aim of the project described here had been to make use of
slew data to help define the EPIC background and vignetting. As no
significant amount of useful slew data has so far been acquired, it
was decided to use a large number of XMM-Newton pointed observations
to produce background maps for each of the three EPIC instruments (pn,
MOS1 \& MOS2) and in several different energy bands. Also,
significantly improved background event files, extremely useful in
terms of spectral analysis, with longer exposures than previosuly
produced files, and specific to several particular
instrument/mode/filter combinations, have been created as a by-product
of the analysis. The adding together of many fields allows the
minimization of any `cosmic variance', resulting from variations in
the local diffuse X-ray emission or contamination from pathologically
bright sources.

This document is intended as an aid when using these files. The
current understanding of the XMM-Newton background is described
briefly in Section 2. Section 3 describes the methods used in the
creation of the background products. Information regarding how to use
the background images for XMM EPIC background analysis can be found in
Section 4, along with caveats as to their use. The relevant
information for the event files can be found in Section 5.

All the background product files (maps, event files, related software
\etc) are available together with this document, and other scripts and
procedures to do with XMM-Newton Background Analysis on
\verb£http://www.sr.bham.ac.uk/xmm3/£

\section{The XMM-Newton X-ray Background} 

The EPIC background has been shown (via the work of Lumb \etal\ (2002)
and many others; see Appendix) to mainly comprise:

\begin{itemize}

\item Solar soft protons (see Sect.3.2) perhaps accelerated by
`magnetospheric reconnection' events, and gathered by XMM-Newton's
grazing mirrors (and perhaps trapped beforehand by the Earth's
magnetosphere). These dominate the times of high background. During
quiescent periods (\ie with no significant soft proton contamination),
the remaining components are:

\item High energy, non-vignetted cosmic ray induced events, unrejected
by the the on-board electronics. Also, associated instrumental
fluorescence, due to the interaction of high-energy particles with the
detector.

\item Non-vignetted electronic noise \ie bad (bright) pixels (and dark
current, though this may be negligible).

\item Low to medium energy, vignetted X-ray photons from the sky. This
can be divided into the local (predominately soft) X-ray background,
the cosmic (harder) X-ray background, and single reflections entering
the telecope from bright sources outside of the nominal field of view
(FOV). Lumb \etal\ (2002) estimate that the contribution of diffuse
flux gathered from out-of-field angles of 0.4$-$1.4 degrees is of
order 7\% of the true, focussed in-field signal, and the associated
systematic error (due mainly to the energy dependence) is $\pm$2\%.

\end{itemize}

Table~1 gives a very brief summary of the temporal, spatial and
spectral properties of these, the major components contributing to the
XMM-Newton background.

\begin{sidewaystable}
\caption[]{Summary of the components within the XMM-Newton Background;
temporal, spatial and spectral properties}
\begin{tabular}{|l|l|l|l|l|l|}
\noalign{\smallskip}
\hline
\noalign{\smallskip}
\hline
  & \multicolumn{2}{c|}{PARTICLES} &  & \multicolumn{2}{c|}{PHOTONS} \\ \hline
\noalign{\smallskip}
\hline
  & SOFT PROTONS & INTERNAL             & ELECTRONIC & HARD X-RAYS & SOFT X-RAYS \\
  &              & (Cosmic-ray induced) & NOISE &             &             \\ \hline
\noalign{\smallskip}
Source          & Few 100\,keV          & Interaction of High & 1) Bright pixels           & X-ray background      & Local Bubble \\ 
                & solar protons         & Energy particles    & 2) Electronic overshoot    &         (AGN etc)     & Galactic Disk \\ 
                &                       & with detector       & near pn readout            &                       & Galactic Halo \\ \hline
\noalign{\smallskip}
Variable?        & & & & & \\ 
(per Obs)       & Flares ($>$1000\%)    & $\pm$10\%           & $\pm$10\%  & Constant              & Constant \\
(Obs to Obs)    & Unpredictable. More   & $\pm$10\%           & 1) $>$1000\% (pixels & Constant       & Variation with \\
                & far from apogee.      & No increase after   & come and go, also &                & RA/Dec ($\pm$35\%) \\ 
                & Low-E flares turn     & solar flares        & meteor damage) & & \\ 
                & on before high-E      & & & & \\ \hline
\noalign{\smallskip}
Spatial         & & & & & \\ 
Vignetted?      & Yes (scattered)               & No                    & No         & Yes           & Yes\\ 
Structure?      & Perhaps, unpredictable        & Yes. Detector +       & Yes & No            & No, apart from real \\  
                &                       & construction &       1) Individual pixels            &      & astronom.\,objects \\  
                &                       & MOS: outer 6 CCDs more        & \& columns        &                       & \\  
                &                       & Al, CCD edges more Si         & 2) Near pn     & & \\  
                &                       & PN: Central hole in high-E& readout (CAMEX) &    &  \\ 
                &                       & lines ($\sim$8\,keV) &   &  & \\    \hline
\noalign{\smallskip}
Spectral        & Variable              & Flat + flourescence +        & 1) low-E ($<$300\,eV), & $\sim$1.4 power law.& Thermal with \ltsim 1keV \\  
                & Unpredictable         &  detector noise              & tail may reach higher-E & Above 5keV,         &  emission lines \\  
                & No correlation between        & MOS: 1.5\,keV Al-K   & 2) low-E ($<$300\,eV) & dominates over      & \\  
                &  intensity + shape            &     1.7\,keV Si-K    & & internal component  & \\  
                & Low-E flares turn     &     det.noise$<$0.5\,keV.    & & & \\  
                & on before high-E      &     High-E $-$ low-intensity & & & \\  
                &                       &     lines (Cr, Mn, Fe-K, Au) & & & \\  
                &                       & PN:  1.5\,keV Al-K           & & & \\  
                &                       &     no Si (self-absorbed)    & & & \\  
                &                       &     Cu-Ni-Zn-K ($\sim$8\,keV)& & & \\  
                &                       &     det.noise$<$0.3\,keV     & & & \\  \hline
\noalign{\smallskip}
\hline
\end{tabular}
\end{sidewaystable}

The present analysis is related in several respects to the work of
Lumb \etal\ (2002), and the reader is encouraged to consult this work.
Additional notes regarding other related work on the EPIC background
can be found in the Appendix.

\section{Analysis} 

Source-subtracted, high-background-screened and exposure- and
source-corrected images (maps) of the particle and photon components
of the EPIC background have been created separately for each EPIC
instrument, and in several different energy bands. This has been
performed separately per observation, over a large number of
individual observations.

The individual background maps for a particular instrument and in a
particular energy band have then then been collected together (for the
same instrument mode/filter combination) over the whole set of
observations. Via various cleaning, filtering and `$\sigma$-clipping'
techniques, a `mean' background map is created (for each particular
background component/instrument/energy band/mode/filter combination).

Procedures have been written to perform the different aspects of the
analysis, making extensive use of the XMM-SAS tasks, some Chandra CIAO
tools and HEASOFT's FTOOLS utilities.

Before discussing the analysis in depth, it is worth describing some
of the files and terms used, and defining the structure of some of the
final products.

\subsection{Analysis: Files \& Definitions}

Here are defined a number of terms used in the following description
of the analysis, and of the product files and maps, and their usage.

\begin{itemize} 

\item Background maps. Detector maps of all the background components
combined \ie soft and hard X-ray photons (vignetted), internal
particles (non-vignetted), some small factor of soft protons
(scattered/funelled) and single-reflections from out-of-FOV sources.

\item Vignetting maps. Detector maps of the effective area \ie the
above background maps minus the particle and single-reflection
contributions.

\item Detector maps. Images in detector coordinates (DETX DETY) (as
opposed to sky coordinates), defined in the present analysis such that
the individual pixels are 1\arcm\ by 1\arcm, DETX/DETY of (0,0) lies
at a pixel intersection, and the full area of the CCDs is
covered. More detailed information as to the coordinate system used
(and of the 4\arcs\ detector maps used within the analysis) is given
in Table~2. Software has been written (see Sect.~4.2) to rebin the
detector maps to any spatial scale, and to reproject the maps to any
sky coordinates (given via a user-input sky image).

\begin{table*}
\caption[]{Detailed information as to the coordinate systems of the
1\arcm\ background detector maps and the 4\arcs\ detector maps used
within the analysis. }
\begin{tabular}{cccccccccc}
\noalign{\smallskip}
\hline
\noalign{\smallskip}
\hline
Instrument & Coord. & \multicolumn{2}{c}{Range in        } & \multicolumn{6}{c}{Number of pixels in maps} \\
           &(DETX/Y)& \multicolumn{2}{c}{Detector coords.} & \multicolumn{3}{c}{Small-scale 
(4\arcs)} & \multicolumn{3}{c}{Large-scale (1\arcm)} \\ 
       &        & min. & max.                          & N$_{\mbox -ve}$ & N$_{\mbox +ve}$ 
& N$_{\mbox tot}$ & N$_{\mbox -ve}$ & N$_{\mbox +ve}$ & N$_{\mbox tot}$ \\ \hline

PN & DETX & -19199 & 14400 & 240 & 180 & 420 & 16 & 12 & 28 \\
   & DETY & -16799 & 15600 & 210 & 195 & 405 & 14 & 13 & 27 \\
M1 & DETX & -20399 & 20400 & 255 & 255 & 510 & 17 & 17 & 34 \\
   & DETY & -20399 & 20400 & 255 & 255 & 510 & 17 & 17 & 34 \\
M2 & DETX & -20399 & 20400 & 255 & 255 & 510 & 17 & 17 & 34 \\
   & DETY & -20399 & 20400 & 255 & 255 & 510 & 17 & 17 & 34 \\ \hline
\noalign{\smallskip}
\hline
\end{tabular}
\end{table*}

\item Instruments. The present analysis has been performed for each of
the EPIC instruments; pn, MOS1 \& MOS2 (hereafter PN, M1, M2).

\item Instrument mode \& filter. Several different data acquisition
modes exist for each of the EPIC instruments. Also, observations have
been taken using different filters for each of the EPIC
instruments. The present analysis has been performed for the most
common instrument mode/filter combinations, and those most useful to
the analysis of diffuse X-ray emission from extended objects. These
are full-frame mode with thin filter (ft) and full-frame mode with
medium filter (fm), for each of PN, M1 \& M2, and also (for PN)
full-frame-extended mode with thin filter (et) and full-frame-extended
mode with medium filter (em).

\item Energy bands. The analysis has been performed in the following
XMM-XID \& PPS standard energy bands; energy bands 1$-$5, defined as
follows; 1: 200$-$500\,eV 2: 500$-$2000\,eV 3: 2000$-$4500\,eV 4:
4500$-$7500\,eV 5: 7500$-$12000\,eV, and the full energy band 0:
200$-$12000\,eV.

\item Closed datasets. A few XMM observations have been obtained with
the filter wheel in the `closed' position. No photons reach the CCDs,
so the event files contain only the instrumental and particle
components of the background. Such datasets have been collected
together and processed, and exist on
\verb£ftp://www-station.ias.u-psud.fr/pub/epic/Closed£ (Marty
2002). They are dependent on the instrumental mode, and have been
processed for M1 full-frame, M2 full-frame, PN full-frame and PN
full-frame-extended mode.

\end{itemize}

\subsection{Analysis: Data Preparation and Reduction}

The initial analysis is essentially the same for each observation.

\begin{itemize}

\item For each observation, the relevant Pipeline Processing System
(PPS) products (event lists, source lists, attitude files,
housekeeping files \etc), from the standard analysis carried out at
the SSC, are collected together.

\item For each instrument, region files are created from the PPS
source lists. These are then used to remove all the source events from
each of the relevant event files. A conservative extraction radius of
36\arcs\ is used to remove the sources (for comparison, Lumb \etal\
(2002) used 25\arcs). All the sources are also removed from previously
created mask files (these are required to calculate losses in area due
to source removal). 

\item A visual inspection is made of the data to make sure that there
are no strange features in the field, and to ascertain whether there
are any very bright point sources or large diffuse sources which could
contaminate the background, even after source subtraction.

\item Each of the event files are then filtered for periods of high
background (solar proton flares). Lightcurves are created over the
whole detector for single events in the energy range 10$-$15\,keV and
with FLAG values as defined by \verb£#XMMEA_EM£ or
\verb£#XMMEA_EP£. \ie with the following expression for the XMM-SAS
task evselect (\eg for PN);

\verb£ PI in [10000:15000] && #XMMEA_EP && PATTERN==0 £

Good Time Interval (GTI) files are created from these lightcurves on
the basis of when the count rate falls below 100 (PN) or 35 (M1/M2)
ct/100\,s. The event files are then filtered, keeping the low count
rate time periods. 

\item The event files are then filtered further. Events with energies
below 150\,eV are discarded. For PN, only singles and doubles are
retained, for M1/M2, singles, doubles, triples and quadruples are
retained. Finally, the event lists are filtered using the
\verb£#XMMEA_EM/P£ FLAG expressions, excepting that events from
outside the field of view (out-of-FOV) are kept.

\item In the case of PN, it was necessary to further filter the files
as regards a small number of persistent bad (bright) pixels/columns,
occurring in many (though not all) of the observations. The same
pixels were also found to contaminate the PN closed datasets. These
pixels were removed from all the PN datasets. Given the large-scale
(1\arcm) of the final BG maps, this made a very small difference, as
the loss in area is very small. Nevertheless, this difference was
corrected for in the final maps. The pixels removed were as follows:
CCD1 col.13 \& (56,75), CCD2 (46$-$47,69$-$72), CCD5 col.11, CCD7
col.34, CCD10 col.61, CCD11 (47$-$48,153$-$156) (50,161$-$164). No bad
pixels were removed from any of the MOS datasets (pointed observations
or closed datasets). The event files are now filtered and have had all
sources removed.

\item For each of the three instruments, a small-scale (4\arcs)
non-vignetted exposure map (with dimensions as given in Table 2) is
created. From the source-removed mask file, an area map (4\arcs) is
created, containing zero values at the positions where sources have
been removed, and unity values elsewhere. These two maps are combined
and rebinned to form a large-scale (1\arcm; see Table 2)
`area-times-exposure' map (see Fig.1).

\begin{figure*}
\vspace{6cm}
\includegraphics{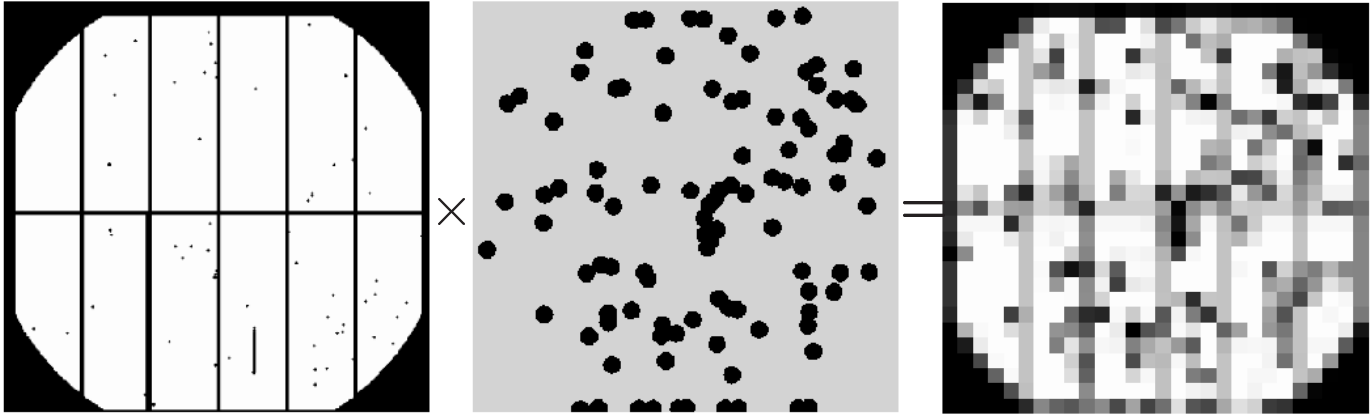}
\caption{A non-vignetted (4\arcs) exposure map is multiplied by a
source-removed (4\arcs) mask map and rebinned to create a large-scale
(1\arcm) `area-times-exposure' map (example is for PN).}
\label{fig1}
\end{figure*}

\item For each instrument (of 3) and each energy range (of 6),
(1\arcm) detector maps are created. Because X-ray photons cannot reach
the out-of-FOV areas of the detector, the events in these areas of the
detector are solely due to the instrumental and particle components of
the background. By making use of the Closed datasets, and comparing
the number of counts in the out-of-FOV regions of the Closed datasets
and the particular observation dataset in question, one can separate
(for each mode-dependent instrument and energy range) the (1\arcm)
detector map into a particle (pa) map, containing the particle and
instrumental background components and a photon (ph) map, containing
the photon component of the background. This is done, producing 2 maps
for each instrument and each energy range. 

\item The particle images are then exposure-corrected via a direct
division by the LIVETIME (corrected for periods of high background and
deadtime, deadtime being times when detector areas are affected by
cosmic rays, and therefore not available to detect valid X-rays), and
a division by a 1\arcm\ detector mask of the sensitive area of the
CCDs (to correct for chip gaps). The photon images are
exposure-corrected via a division by the appropriate
`area-times-exposure' map.

\end{itemize}

\subsection{Analysis: Observation, Mode and Filter Selection}

The above preparational analysis has been performed for a large number
of XMM-Newton observations, spanning a large range in instrumental
modes used, filters used, exposure times and degree and duration of
high-background flaring. 

In order to produce the final background maps, the observations have
been collected together in terms of the above requirements. Further,
all observations containing a significantly bright source whose wings
could still contaminate the background after source subtraction have
been removed from the subsequent analysis. Similarly, all observations
containing a large diffuse source, which could contribute to the
estimated background, have been removed. All observations where the
background flaring was of such an extent that, after flare-removal,
less than 10\% of the original exposure remained, were also removed.

The remaining 72 `clean' observations used in the production of the
final background files are summarized in Table~3, which lists the
revolution number, the source target name, and a code giving the
instrumental mode and filter for, respectively, M1, M2 and PN [f -
full-frame mode, e - full-frame-extended mode , t - thin filter, m -
medium filter]. Also given are the nominal exposure time, and the
fraction of the exposure time remaining after removal of
high-background periods and accounting for deadtime effects.

\begin{sidewaystable}
\vspace{15mm}
\caption[]{Summary of the cleaned and filtered observations used in
the production of the EPIC background files. Given is the revolution
number, the source target name, and a code giving the instrumental
mode and filter for M1, M2 and PN [f - full-frame mode,
e - full-frame-extended mode , t - thin filter, m - medium
filter]. Also given is the nominal exposure time, and the fraction of
the exposure time remaining after removal of high-background periods. }
\begin{tabular}{|clcrc|clcrc|}
\noalign{\smallskip}
\hline
\noalign{\smallskip}
\hline
Rev.& Source & Mode/Filter & Expos. & f(Exp.) & Rev.& Source &
    Mode/Filter & Expos. & f(Exp.) \\ & & (M1M2PN) & (s) & & & &
    (M1M2PN) & (s) & \\ \hline

281 & MS1054.4-0321 	& ft\,\,ft\,\,et\,\, & 40260 & 0.61 & 324 & PMN0525-3343 	& ft\,\,ft\,\,ft\,\, & 27786 & 0.65 \\ 
283 & IC1623 		& fmfmfm & 11813 & 0.13 & 339 & NGC2563 	& fmfmfm & 21617 & 0.88 \\ 
284 & C2001 		& ft\,\,ft\,\,ft\,\, &  9900 & 0.40 & 343 & S308 		& fmfmem & 47559 & 0.37 \\ 
285 & ESO263-6048 	& fmfmem &  9215 & 0.89 & 344 & IRAS07598+6508 	& fmfmfm & 22615 & 0.73 \\ 
286 & NGP rift1 	& ft\,\,fmet\, & 24211 & 0.81 & 345 & SC223 		& fmfmem & 34326 & 0.45 \\ 
286 & NGP rift2 	& ft\,\,fmet\, & 24211 & 0.89 & 345 & 3C192 		& ft\,\,ft\,\,et\, & 24326 & 0.70 \\ 
286 & NGP rift3 	& ft\,\,fmet\, & 24214 & 0.66 & 346 & 3EG0616-3310 	& ft\,\,ft\,\,et\, & 11773 & 0.90 \\
287 & XMDSOM 5 		& ft\,\,ft\,\,et\, & 24218 & 0.89 & 346 & 3EG0616-3310 	& ft\,\,ft\,\,et\, & 14772 & 0.46 \\ 
287 & XMDS SSC 3 	& ft\,\,ft\,\,et\, & 24318 & 0.89 & 346 & 3EG0616-3310 	& ft\,\,ft\,\,et\, & 14771 & 0.71 \\ 
287 & XMDS SSC 4 	& ft\,\,ft\,\,et\, & 28512 & 0.88 & 346 & APM08279+5255 	& fmfmfm & 16916 & 0.90 \\ 
287 & XMDS SSC 5 	& ft\,\,ft\,\,et\, & 24313 & 0.87 & 347 & CEGru 		& ft\,\,ft\,\,ft\,\, &  7615 & 0.91 \\ 
288 & 1saxj2331.9+1938 	& fmfmfm & 11513 & 0.77 & 348 & NGC3184 	& ft\,\,ft\,\,ft\,\, & 29617 & 0.76 \\ 
288 & XMDS SSC 1 	& ft\,\,ft\,\,et\, & 25149 & 0.74 & 348 & RXJ0911.4+0551 	& ft\,\,ft\,\,et\, & 20072 & 0.22 \\ 
288 & XMDS SSC 2 	& ft\,\,ft\,\,et\, & 12268 & 0.17 & 349 & Lockman Hole 	& fmfmfm & 37950 & 0.77 \\ 
288 & SGP-3 		& ft\,\,ft\,\,et\, &  9209 & 0.62 & 353 & Arp270 		& ft\,\,ft\,\,et\, & 37800 & 0.36 \\ 
299 & AXAF Ultra Deep 	& ft\,\,ft\,\,et\, & 39765 & 0.49 & 353 & RXJ1011.2+5545  & xx\,\,ft\,\,ft\,\, & 32615 & 0.83 \\ 
299 & AXAF Ultra Deep 	& ft\,\,ft\,\,et\, & 56698 & 0.65 & 353 & UGC05101 	& fmfmet\, & 34371 & 0.89 \\ 
300 & H1413+117 	& ft\,\,ft\,\,et\, & 25719 & 0.88 & 354 & NGC7252 	& fmfmfm & 27617 & 0.85 \\ 
305 & ERO field 	& ft\,\,ft\,\,ft\,\, & 40387 & 0.86 & 354 & MS2053.7-0449 	& xx\,\,fmem & 16770 & 0.89 \\ 
307 & Omega cen 	& fmfmfm & 39978 & 0.47 & 355 & LBQS2212-1759 	& ft\,\,ft\,\,xx\,\, & 34397 & 0.96 \\ 
308 & XMDS SSC 2 	& ft\,\,ft\,\,et\, & 12666 & 0.89 & 355 & LBQS2212-1759 	& ft\,\,ft\,\,ft\,\, & 10824 & 0.62 \\ 
310 & XTEJ0421+560 	& fmfmfm & 32310 & 0.40 & 356 & LBQS2212-1759 	& ft\,\,ft\,\,ft\,\, &109427 & 0.77 \\ 
311 & CenX-4 		& ft\,\,ft\,\,ft\,\, & 52799 & 0.43 & 359 & NGC3044 	& ft\,\,ft\,\,et\, & 10020 & 0.39 \\ 
312 & RXJ0002+6246 	& fmfmsm & 33000 & 0.49 & 360 & SextansA 	& fmfmem & 19130 & 0.21 \\ 
312 & S50014+813 	& fmfmfm & 39337 & 0.35 & 360 & PG1115+080 	& fmfmft\,\, & 62616 & 0.78 \\ 
314 & NGC2146 		& fmfmft\,\, & 26847 & 0.40 & 360 & NGC4138 	& ft\,\,ft\,\,em & 14317 & 0.87 \\ 
316 & GRS1716-249 	& ft\,\,ft\,\,ft\,\, & 12225 & 0.70 & 361 & Arp222 		& ft\,\,ft\,\,et\, & 14369 & 0.86 \\ 
316 & GROJ1655-40 	& fmfmfm & 39820 & 0.46 & 361 & RXJ2237.0-1516 	& ft\,\,ft\,\,et\, & 24370 & 0.78 \\ 
317 & CFHT-PL-12 	& ft\,\,ft\,\,ft\,\, & 34335 & 0.31 & 362 & PG2302+029 	& ft\,\,ft\,\,ft\,\, & 12566 & 0.88 \\ 
317 & L1448-C 		& ft\,\,ft\,\,ft\,\, & 32725 & 0.32 & 363 & LSBC F568-6 	& ft\,\,ft\,\,et\, & 14365 & 0.47 \\ 
317 & SHARC-2 		& ft\,\,ft\,\,et\, & 48870 & 0.64 & 364 & NGC4168 	& ft\,\,ft\,\,em & 22867 & 0.86 \\ 
321 & oph pos1 		& ft\,\,ft\,\,et\, & 20121 & 0.88 & 366 & CSO755 		& ft\,\,ft\,\,ft\,\, & 36926 & 0.72 \\ 
321 & oph pos2 		& ft\,\,fmet\, & 18626 & 0.82 & 383 & LSS 2 		& ft\,\,ft\,\,et\, & 13382 & 0.85 \\ 
322 & VIIZw031 		& fmfmfm & 31997 & 0.90 & 383 & LSS 6	 	& ft\,\,ft\,\,et\, & 13382 & 0.44 \\ 
323 & KS1731-260 	& ft\,\,ft\,\,ft\,\, & 24618 & 0.61 & 383 & LSS 7 		& ft\,\,ft\,\,et\, & 12380 & 0.87 \\ 
323 & KS1731-260 	& ft\,\,ft\,\,ft\,\, & 25115 & 0.60 & 451 & NGC4490 	& fmfmem & 17521 & 0.85 \\ \hline
\noalign{\smallskip}
\hline
\end{tabular}
\end{sidewaystable}

Table~4 summarizes the final cleaned observation information in terms
of the different combinations of instrument, instrument mode and
filter used.  The exposure time is the sum of the individual LIVETIMEs
\ie corrected for periods of high-background and deadtime.

\begin{table}
\caption[]{Summary of the cleaned and filtered observations used in
the production of the EPIC background files, separated into the
different combinations of instrument, instrument mode and filter
used. The exposure time is the sum of the individual LIVETIMEs \ie
corrected for periods of high-background and deadtime.}
\begin{tabular}{lllrr}
\noalign{\smallskip}
\hline
\noalign{\smallskip}
\hline
Instrument & Mode                & Filter & Number of    & Exposure   \\
           &                     &        & Observations &   Time (s) \\ \hline
 MOS1      & Full-Frame          & Thin   & 49           &    1055905 \\
 MOS1      & Full-Frame          & Medium & 21           &     488422 \\
 MOS2      & Full-Frame          & Thin   & 46           &    1004709 \\
 MOS2      & Full-Frame          & Medium & 26           &     592975 \\
 PN        & Full-Frame          & Thin   & 18           &     351549 \\
 PN        & Full-Frame          & Medium & 12           &     188159 \\
 PN        & Full-Frame-Extended & Thin   & 32           &     416739 \\
 PN        & Full-Frame-Extended & Medium &  8           &      82957 \\ \hline
\noalign{\smallskip}
\hline
\end{tabular}
\end{table}

\subsection{Analysis: Cubes and Clipping}

The basic principle involved in the production of a single `averaged'
background map from many background maps over different observations
is that of removing the outliers which may represent images that are
contaminated at a particular pixel.

In order to produce a particular set of background maps (in the
different energy bands) for a particular instrument/mode/filter
combination, the following steps are taken. Let us take for example
the first entry in table~4, \ie MOS1, full-frame mode, thin filter.

For each of the 49 relevant observations, comprising in total over
1\,Msec of clean, low-background data, the 0-band (full energy band)
1\arcm\ photon detector maps of the background are stacked together
into a 3-D `imagecube' (see Fig.\,2) of dimensions DETX, DETY (each in
the case of MOS1 of size 34; see Table~2) and N$_{\mbox {\small obs}}$ (of
size, in this example, 49).

\begin{figure*}
\vspace{9cm}
\includegraphics{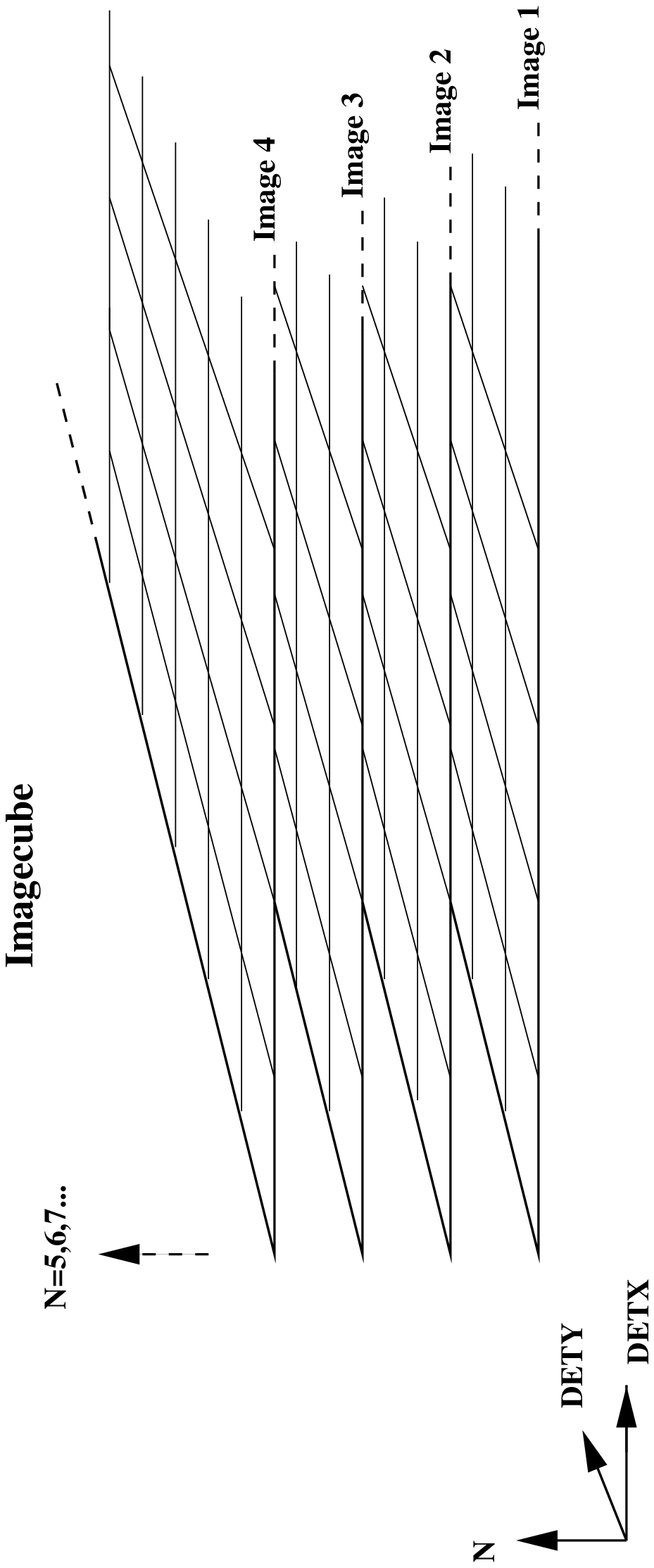}
\caption{A 3-D `imagecube' of dimensions DETX, DETY and N$_{\mbox
{\small obs}}$.}
\label{fig2}
\end{figure*}

For each DETX/DETY, a statistical analysis of the (in this case, 49)
values at that particular DETX/DETY point is performed. Here, a
`clipcube' is constructed (of dimensions identical to the input
imagecube), containing information (1's and 0's) as to which cells in
the imagecube are within the allowed range and which values are not,
\ie which values are `clipped' (see Fig.~3). The allowable range is
defined as within some number of standard deviations ($\sigma$s) from
the mean value at that DETX/DETY. For the present analysis, this
number of $\sigma$s is set at 1.2. Several steps are taken to ensure
that the initial mean used to define the clip limits is not seriously
biased by outliers. The mean value at that particular DETX/DETY is
then calculated from the remaining `$\sigma$-clipped' values.

\begin{figure*}
\vspace{9cm}
\includegraphics{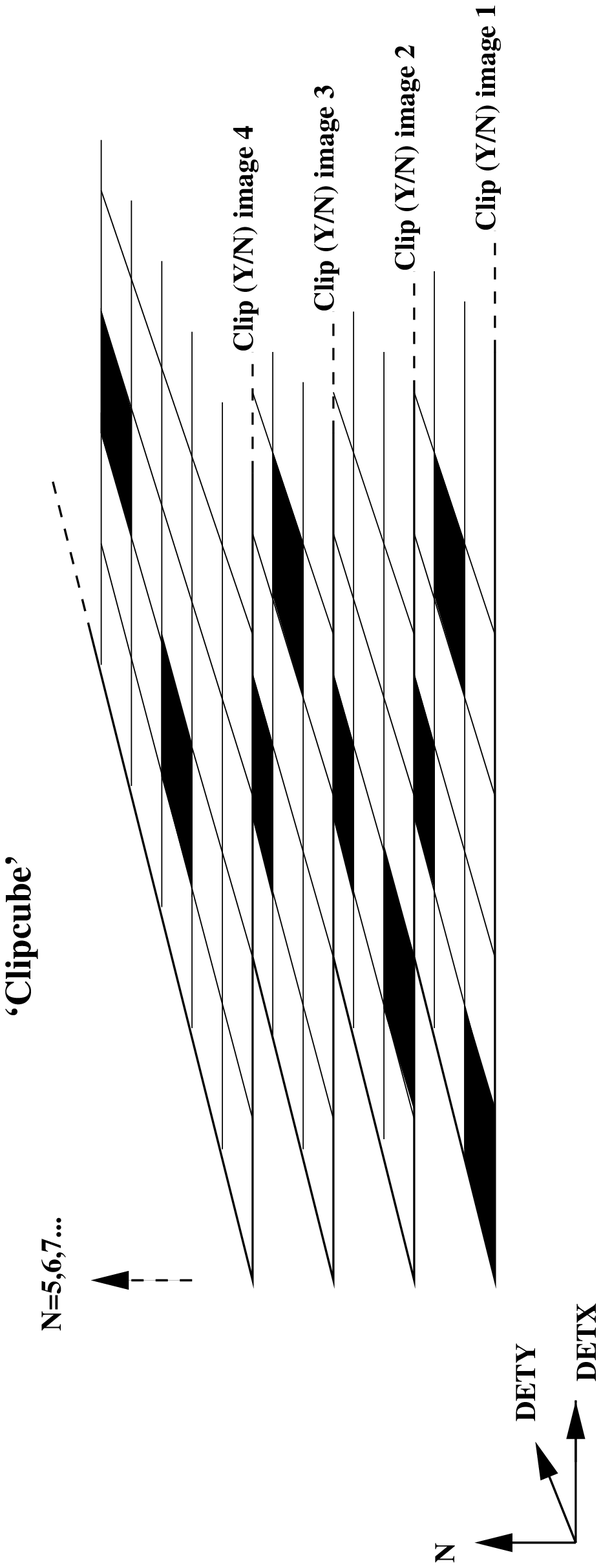}
\caption{A 3-D `clipcube' of dimensions DETX, DETY and N$_{\mbox
{\small obs}}$. In this example, for DETX/DETY=(1,1), the value in
image 1 is outside of the allowable range, and is to be clipped from
the calculation of the mean value for DETX/DETY=(1,1).}
\label{fig3}
\end{figure*}

Imagecubes are then created, as in the 0-band, from the intermediate
energy band (bands 1$-$5) 1\arcm\ photon detector maps of the
background. A $\sigma$-clipped image is then created from these cubes,
but using the previously created relevant 0-band clipcube to remove
values from each (band 1$-$5) imagecube. The same procedure is
performed for the particle images, and for all instrument/mode/filter
combinations. 

\section{The Background maps}

$\sigma$-clipped, and averaged, exposure-corrected background maps
have been created as described above for each instrument/mode/filter
combination analysed (of 8) and in each energy band (of 6). Photon and
particle maps have been created separately, as have maps with the two
components recombined. This leads to a total of 144 background maps. 

\verb£A1_ft0000_cphim4M1.fits£ is, as an example of the file naming,
an exposure-corrected photon background map. A particle map has
\verb£cpaim£, instead of \verb£cphim£, and a particle and photon
combined map has \verb£cim£. It is a MOS1 map (\verb£M1£ instead of
\verb£M2£ or \verb£PN£), and is in energy band \verb£4£ (of
\verb£0£$-$\verb£6£). The mode is full-frame (given by the \verb£f£),
as opposed to (for PN) extended full-frame mode (\verb£e£), and the
filter is thin (\verb£t£), as opposed to medium (\verb£m£). The six
character mode+filter string is as in Table~3, hence the corresponding
PN file has \verb£0000ft£.

\subsection{The Background maps: Usage}

How then can an observer with their own EPIC data make use of these
background maps? In order to analyse extended objects, and make
background-subtracted images, estimate low-surface brightness flux
levels, create radial profiles or perform 1- or 2-D surface brightness
fitting, one needs a map of the appropriate background. Sometimes
however, the diffuse, extended nature of the user's target source is
such that the determination of the background from their own dataset
is difficult or impossible. Also, using a background from a
significantly removed section of the same data leads to problems with
vignetting and detector variations. Hence the need for the
independently-produced background maps created here.
 
What follows is a recipe for how an observer can make use of the
background maps in conjunction with their own source data. Caveats,
and problems that can occur when the recipe is not or can not be
followed are given thereafter.

\begin{itemize}

\item The source data should be flare-rejected to a similar level to
that performed in the creation of the background maps. $^1$

\item Source maps should be created from the flare-rejected source
datasets in the same XID/PPS energy bands (bands 0$-$5) as used
here. $^2$

\item As a particular source map will be created (most commonly) in
sky coordinates, and will have a resolution finer than 1\arcm, the
routine {\em BGrebinimage2SKY} should be used to rebin and sky-project
the equivalent background map(s) to the resolution and sky position of
the source map (see Sect.\,4.2). 

\item Depending on whether the source maps are raw count maps or
exposure-corrected flux maps, the equivalent rebinned background
map(s) (which are exposure-corrected) may be scaled appropriately,
either by an exposure time, or using an appropriate exposure map.

\item A comparison of the counts or flux in the out-of-FOV areas of
the source and background maps yields the scaling by which the particle
component (the `cpaim' map) of the background needs be scaled. $^3$

\item If source-free regions within the FOV exist within the source
dataset, then count/flux comparisons in these regions yield the
scaling for the photon component (the `cphim' map). The scaled photon
component can then be add to the scaled particle component to give the
final background map. $^3$

\end{itemize}

$^1$ If such a flare-rejection method as used here leads to zero or
very few Good Times, then the user's data is heavily
flare-contaminated, and the background maps are therefore not suitable
for source data extracted from the whole dataset. Small discrepancies
due to slightly different flare-rejection methods may be accounted for
by the subsequent scaling.

$^2$ Though the user should work using the same energy bands (bands
0$-$5) as used in the present analysis, note that the standard XID/PPS
bands have been used, and that extension to larger energy bands can
easily be performed by summing individual band images. For example,
band 1 is heavily contaminated by detector noise in the particle
background, so the user may prefer to work in band (2+3+4+5).

$^3$ If source-free regions within the FOV do exist within the source
dataset, then the user is advised to work as above with the particle
and photon (`cpaim' \& `cphim') maps separately. The particle and
photon scaling factors may be different (hence the fact that separate
particle and photon background maps have been made available). If no
source-free regions exist, the user may be forced to assume that the
scalings are the same. Here, just the `cim' images need be used. A
script for comparing out-of-FOV counts (from event files), {\em
compareoutofFOV}, is available on http://www.sr.bham.ac.uk/xmm3/. Note
especially that at the very edges of the map FOVs, there are a few
pixels with unusually large and small values. This is due to extremely
small exposure values in the original observation maps amplifying the
noise. These areas should be avoided when calculating scaling values.

Note also that comparisons of the in- and out-of-FOV source maps are
useful in characterizing the low-energy flare contamination (\eg de
Luca \& Molendi 2002).

\subsection{The Background maps: Sky coordinates, Rebinning and Reprojecting}

Usually, of course, the user will work in sky coordinates (RA \& Dec),
not detector coordinates, and to a much finer scale than 1\arcm. With
that in mind, software has been developed to rebin and reproject onto
the sky any provided background map (scaled or otherwise) to the
spatial scale and sky position of a user-input image.

BGrebinimage2SKY is a shell script plus fortran routine to convert the
low-resolution DETX/DETY background maps into high-resolution sky
(X/Y) images. The user gives a template image containing the attitude
information, and an event file (ideally, the one used to create the
template image file, though any may do, as it is just for general
header purposes). A rebinned background map is produced with the same
resolution and at the same sky position as the template image. The
radius of the interpolation circle can be given (note that the
background DETX/DETY maps are of 1\arcm\ resolution, so an
interpolation radius of this or slightly larger is recommended).

\verb£Use: BGrebinimage2SKY image imtemplate evfile rinterp outimage £

\verb£  image      - Input image (low-res DETX/DETY image) £ 

\verb£  imtemplate - Template image file containing attitude info (high-res SKY image) £ 

\verb£  evfile     - Event file, ideally the one used to create imtemplate £ 

\verb£  rinterp    - Interpolation radius [arcmin] (1 - 1.5) £ 

\verb£  outimage   - Output image (high-res SKY image) £ 

\verb£e.g. BGrebinimage2SKY BGimage.fits myimage.fits evfile.fits 1.2 newBGimage.fits £

An example of the task in use can be seen in Fig.4. 

\begin{figure*}
\vspace{17cm} \includegraphics{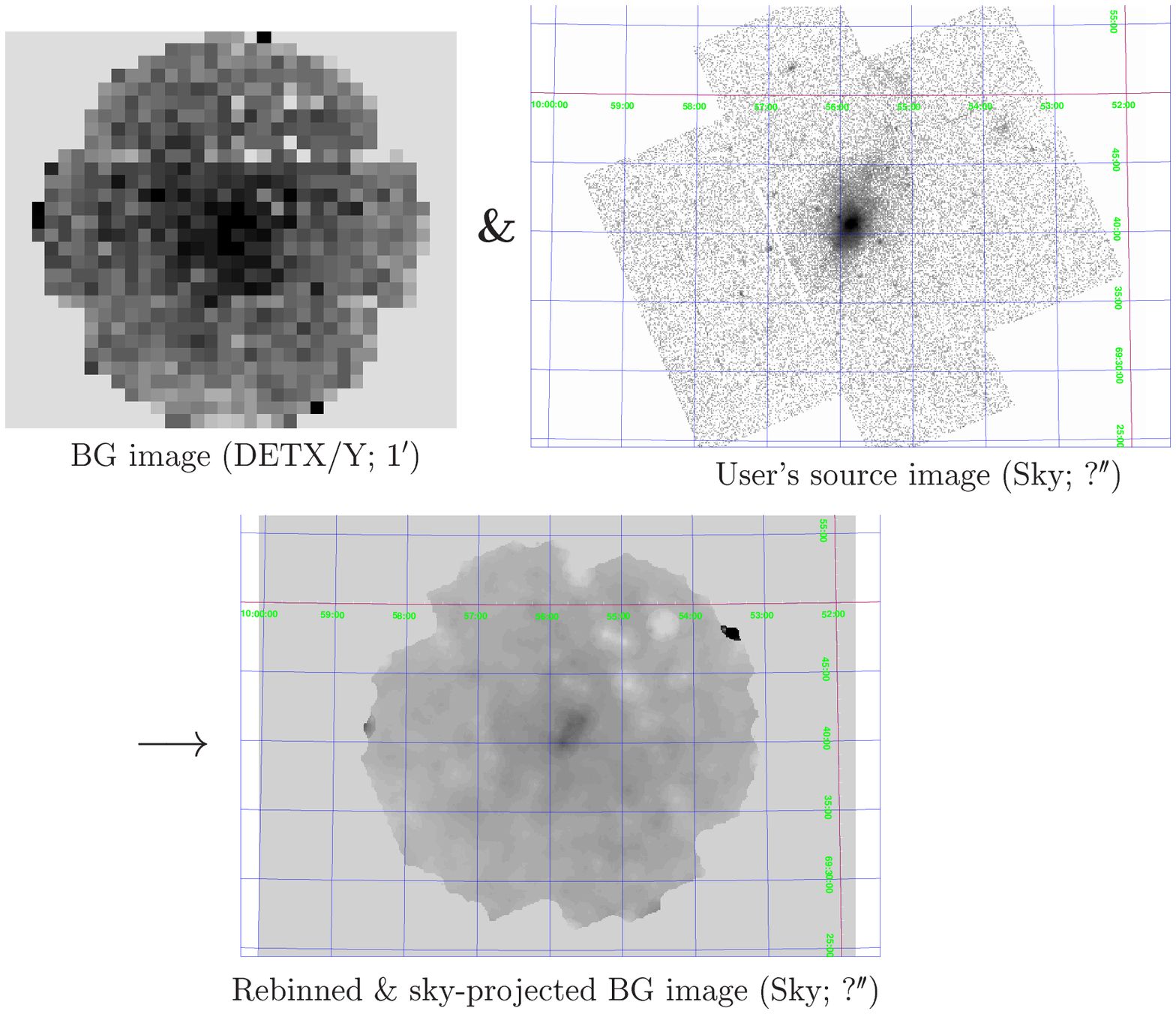}
\caption{Usage of BGrebinimage2SKY to rebin and sky-project a given
low-resolution DETX/Y background map (top left) to the spatial binning
and sky position of a user-created high-resolution sky image (top
right). The resultant high-resolution sky background image is shown at
the bottom. (Note that a poor quality, immature version of a
background image has been used to show more clearly the rebinning and
reprojecting.}
\label{fig1}
\end{figure*}

\section{Event files}

As a by-product of the analysis, for each observation, event files have
been created, in the same manner as Lumb \etal\ (2002), filtered for
times of high background and with all sources removed. The relevant
event files have been merged together for each instrument/mode/filter
combination (Tables 3 \& 4), and eight separate event files have
therefore been created (with exposure times given as in Table 4), each
having an extension containing a calibrated event list in the same
format as produced by the XMM-SAS. The event files have had sky
coordinates assigned to them for a pointing on the sky of RA=0, Dec=0,
PA=0. These event files (or indeed any) can be reprojected onto any
point in the sky via \eg {\em skycast} (see
http://www.sr.bham.ac.uk/xmm3/).

It is believed that these event files offer several improvements over
previous versions (\eg\ those of Lumb \etal\ (2002), themselves
improvements on previously-created versions) for several
reasons. Important points are as follows (note that many of the points
are particularly important not only as regards the event files, but
also the above background maps);

\begin{itemize}

\item The event files have been created separately for each
combination of instrument, instrument mode and filter. Hence, event
files now exist for medium filter and also for pn full-frame-extended
mode, neither of which had existed previously.

\item As stated in the analysis sections, all data are collected from
source-subtracted and high-background filtered fields with no bright
sources or diffuse features, which could contaminate the `background',
even after source-subtraction.

\item The datasets are very long, longer than previously created event
files (over a million seconds of clean low-background data exists in
each of the thin filter MOS datasets, for example). This allows
further improved signal-to-noise.

\item The secondary extensions of the files have been removed, and all
headers cleaned and corrected. This results in far smaller and more
manageable files. It is possible to treat the event files as normal
SAS event files; images, spectra and lightcurves can be created via
evselect, and one can run ftools on the files.

\item The event files consist of many more shorter exposure
observations (rather than a few, long observations), such that the
holes left by the source-subtraction are of less importance, and are
heavily diluted by data from other observations. Due partly to the
different exposure times of the individual observations, no definite
flux limit exists above which sources have been removed. However, flux
histograms of the detected and thereafter removed sources do show
quite a sharp cutoff at $\approx 1\times10^{-14}$\,erg cm$^{-2}$
s$^{-1}$) (see \verb£http://www.sr.bham.ac.uk/xmm3/BGproducts.html£).

\item The exposure times are believed to be accurate, and include
losses due to deadtime.

\end{itemize}

\subsection{Event files: usage and caveats}

The eight event files are named in a similar manner to the background
maps. \verb£E1_ft0000_M1.fits£ is a MOS1 full-frame, thin filter event
file. \verb£PN£ and \verb£M2£ are for the pn and MOS2 instruments, an
\verb£e£ denotes extended full-frame mode, and an \verb£m£, medium
filter.

As regards using the event files for background analysis, it is
strongly suggested consulting Lumb \etal\ (2002). Because of the above
improvements, a number of the caveats are now not relevant as regards
the present files. Many however, are still valid.

\begin{itemize}

\item The user should take care in extracting background from an
appropriate region. This can be done in detector coordinates, though
the background event files can be {\em skycast} onto the sky position
of the user's field (see http://www.sr.bham.ac.uk/xmm3/ -
alternatively use the XMMSAS task {\em attcalc}).

\item Remnant low-level flares will remain in the event files. It is
known, for instance, that at low energies, the proton flares turn on
more slowly, but earlier that the main flare. The user is able to
apply further, more stringent flare-screening to the event files than
that performed in their formation, but note that the lowest level
proton fluxes may be spectrally variable, so that no complete
subtraction may be possible.

\item Although point sources have been removed, examination of images
created from the datasets does reveal fluctuations in intensity. Over
scales of arcminutes appropriate to extended sources, it is not
expected to be a significant problem, and indeed representative of
unresolved background. If a user however, should try to extract
spectra from regions comparable to the PSF scale, care should be
taken, and a manual inspection may be necessary to guard against a
local excess or deficit of counts arising from the point source
extraction procedures. In some of the present cases, it is true that,
as many of the observations were pointed such that the target source
was at the centre of the detector, a deficit in counts is seen at this
position.

\item Many defects are seen at the lowest energies (below 0.3\,keV),
and the calibration of the EPIC response at these energies is not so
well understood as it is at higher energies. Care should be taken when
performing analysis below around 0.25\,keV.

\end{itemize}

\section{Final remarks}

At this time of writing, there has been only scant experience gained
in using either the maps, the event files or the related software. Any
feedback on using the datasets would be very useful and is most
welcome (please email any comments to \verb£amr@star.sr.bham.ac.uk£).

It is hoped that further releases of the datasets, created using
larger numbers of pointed observations, and perhaps using further
modes and filters, will be made available in the future. This will be
announced in the usual manner and via the URL
\verb£http://www.sr.bham.ac.uk/xmm3/£.

AMR wishes to acknowledge Trevor Ponman \& Laurence Jones (Birmingham)
for very useful discussions during this work, and Mike Denby
(Leicester, SSC) for making many datasets available.


\section*{Appendix: The XMM-Newton X-ray Background $-$ Additional notes}

Several members of the community have been analysing various aspects
of the XMM-Newton background, and many of these are described in Marty
(2002), Lumb (2002) \& Lumb \etal\ (2002). The reader is strongly
encouraged to study Lumb \etal\ (2002).

A number of datasets have also been analysed, collected and merged
together. Two sets of data are particularly relevant as regards the
present analysis:

Background datasets have been produced by Dave Lumb (Lumb 2002,
Lumb \etal\ 2002) for the three EPIC instruments by
source-subtracting and co-adding a few long observations. These
event files can be obtained from
\verb£http://xmm.vilspa.esa.es/ccf/epic/#background£, along with very
useful explanatory notes
(\verb£http://xmm.vilspa.esa.es/docs/documents/CAL-TN-0016-2-0.ps.gz£)

Event lists combining several CLOSED observations have been created by
Phillipe Marty, and these have also proven very useful in the analysis
and background-subtraction of extended objects (see Sect.3.2). These
data can be obtained from \verb£ftp://www-station.ias.u-psud.fr/pub/epic/Closed£

In addition, several novel methods have been used to analyse very
extended and diffuse X-ray sources, where the extraction of the
background is difficult. Many of these are described in Marty \etal\
2002. Of special interest is the work of Arnaud \etal\ (2001), de Luca
\& Molendi (2002), Ghizzardi \etal\ (2000), Molendi (2001), Pratt
\etal\ (2001) and many others.

\end{document}